\pdfoutput=1
\documentclass[aps,prl,nofootinbib,preprintnumbers,amsmath,amssymb,latexsym,array,enumerate,letter,twocolumn,superscriptaddress]{revtex4}
\usepackage{amsmath,amssymb,amsthm,array,braket,breakurl,color,enumitem,epsfig,feynmf,graphicx,hyperref,mathtools,siunitx,slashed,xcolor}
\usepackage[normalem]{ulem}

\makeatletter

\begin{document}

\preprint{DESY-22-172, IFT-UAM/CSIC-22-140, MITP-22-095, RESCEU-22/22}

\title{Gravitational Waves from Feebly Interacting Particles
\\
in a First Order Phase Transition}

\author{Ryusuke Jinno}
\affiliation{Instituto de F\'{\i}sica Te\'orica UAM/CSIC, C/ Nicol\'as Cabrera 13-15, Campus de Cantoblanco, 28049, Madrid, Spain}
\affiliation{Research Center for the Early Universe, The University of Tokyo, Hongo 7-3-1 Bunkyo-ku, Tokyo 113-0033, Japan}

\author{Bibhushan Shakya}
\affiliation{Deutsches Elektronen-Synchrotron DESY, Notkestr. 85, 22607 Hamburg, Germany}

\author{Jorinde van de Vis}
\affiliation{Deutsches Elektronen-Synchrotron DESY, Notkestr. 85, 22607 Hamburg, Germany}
\affiliation{Institute for Theoretical Physics, Utrecht University, Princetonplein 5, 3584 CC Utrecht, The Netherlands}

\def\be{\begin{equation}}
\def\ee{\end{equation}}
\newcommand{\bea}{\begin{eqnarray}}
\newcommand{\eea}{\end{eqnarray}}

\begin{abstract}
First order phase transitions are well-motivated and extensively studied sources of gravitational waves (GWs) from the early Universe. The vacuum energy released during such transitions is assumed to be transferred primarily either to the expanding bubble walls, whose collisions source GWs, or to the surrounding  plasma, producing sound waves and turbulence, which source GWs. In this Letter, we study an alternative possibility that has not yet been considered: the released energy gets transferred primarily to feebly interacting particles that do not form a coherent interacting plasma but simply free-stream individually. We develop the formalism to study the production of GWs from such configurations, and demonstrate that such GW signals have qualitatively distinct characteristics compared to conventional sources and are potentially observable with near-future GW detectors. 
\end{abstract}

\maketitle

\section{Motivation}

Gravitational waves (GWs) provide a unique probe of a variety of very early Universe phenomena. One of the most attractive targets for GW searches is a first order phase transition (FOPT)~\cite{Hogan:1983ixn,Witten:1984rs,Hogan:1986qda,Kosowsky:1991ua,Kosowsky:1992rz,Kosowsky:1992vn,Kamionkowski:1993fg}, where the metastable early Universe false vacuum decays through nucleation, expansion, and percolation of bubbles of  true vacuum. The properties of GW signals generated by FOPTs have been extensively studied (see e.g.\cite{Caprini:2015zlo,Caprini:2018mtu,Caprini:2019egz,LISACosmologyWorkingGroup:2022jok} for reviews). FOPTs can generically occur in several beyond the Standard Model (BSM) scenarios, where the existence and breaking of additional symmetries in extended sectors (which could include dark sectors) is motivated by various shortcomings of the Standard Model (SM). 
Such dark sector FOPTs can be realized across a broader range of energy scales~\cite{Schwaller:2015tja,Jaeckel:2016jlh,Dev:2016feu,Baldes:2017rcu,Tsumura:2017knk,Okada:2018xdh,Croon:2018erz,Baldes:2018emh,Prokopec:2018tnq,Bai:2018dxf,Breitbach:2018ddu, Fairbairn:2019xog, Helmboldt:2019pan,Ertas:2021xeh}, offering detection prospects with various current and near future GW detectors such as LIGO-Virgo~\cite{LIGOScientific:2016aoc,LIGOScientific:2016sjg}, LISA~\cite{2017arXiv170200786A}, DECIGO~\cite{Kawamura:2006up}, Big Bang Observer (BBO)~\cite{Harry:2006fi}, Einstein Telescope (ET)~\cite{Punturo:2010zz}, and Cosmic Explorer (CE)~\cite{Reitze:2019iox}. 

FOPTs produce GWs in several ways. If the bubble walls carry most of the energy released in the transition, GWs are sourced by the scalar field energy in the bubble walls when the walls collide~\cite{Kosowsky:1991ua,Kosowsky:1992rz,Kosowsky:1992vn,Kamionkowski:1993fg,Huber:2008hg,Bodeker:2009qy,Jinno:2016vai,Jinno:2017fby,Konstandin:2017sat,Cutting:2018tjt,Cutting:2020nla}, or by particles produced from bubble collisions \cite{Inomata:2024rkt}. In the presence of significant interactions between the walls and the plasma, the released energy is instead primarily transferred to the plasma, and GWs are produced by sound waves (SWs)~\cite{Hindmarsh:2013xza,Hindmarsh:2015qta,Hindmarsh:2017gnf,Cutting:2019zws,Hindmarsh:2016lnk,Hindmarsh:2019phv} and turbulence~\cite{Kamionkowski:1993fg,Caprini:2009yp,Brandenburg:2017neh,Cutting:2019zws,RoperPol:2019wvy,Dahl:2021wyk,Auclair:2022jod}. These contributions have distinct spectral features determined by the behavior of the walls or SWs during percolation. 

In this Letter, we study a new source of GWs from FOPTs that has so far not been considered: the energy released in the phase transition (PT) can be transferred primarily to feebly-interacting particles (FIPs) that free-stream without interacting over the timescale of the PT. Such scenarios can readily occur in dark sectors, which can contain particles with feeble interactions in many realistic scenarios \cite{Agrawal:2021dbo}. In such cases, the standard sources mentioned above carry negligible fractions of the total energy, and cannot be efficient GW sources. This seemingly nightmare scenario for GW searches, where a FOPT does not lead to observable signals even with otherwise favorable parameters, deserves greater scrutiny. In this paper, we develop the formalism to study the evolution of such FIPs during the PT, the subsequent production of GWs, and observation prospects with the next generation of GW detectors.

\section{Framework}\label{sec:framework}

Consider a FOPT involving a dark sector scalar $s$ obtaining a vacuum expectation value $\langle s \rangle$ at temperature $T$ (in general, $T\lesssim \langle s \rangle$), producing bubbles of true vacuum (broken phase), whose walls expand into the false vacuum (symmetric phase) with velocity $v_w$ (and Lorentz factor $\gamma_w$).  For simplicity, we assume that the dark sector constitutes the dominant form of radiation, and the energy in the SM bath is negligible.\,\footnote{Including the SM bath will not change any of our discussions qualitatively, but simply dilute the GW signal.} We parametrize the energy density released during the PT as\,\footnote{Strictly speaking, the PT strength should be parameterized by the trace of the energy-momentum tensor, see \cite{Giese:2020rtr,Giese:2020znk}.}
\be
\alpha=\frac{\Delta V}{\rho_{\rm rad}}=\frac{\Delta V}{\frac{\pi^2}{30}\, g_*^DT^4}\, ,
\ee
where $\Delta V$ is the difference in the potential energies of the two vacua, $\rho_{\rm rad}$ is the radiation energy density, and $g_*^D\!=\!g_{\rm bosons}+(7/8) g_{\rm fermions}$ represents the total number of degrees of freedom in the dark sector. 
$\beta/H$ denotes the inverse timescale for the phase transition normalized by the Hubble time; for $v_w\approx 1$, $\beta$ also represents the average bubble size at collision. 

Consider a particle $X$ in the bath that is massless in the false vacuum but obtains a mass $m$ in the true vacuum due to its coupling to $s$. A massless $X$ particle with energy $E$ can cross into the bubble wall only if $\gamma_w\, E \gtrsim m$. 
The pressure on the bubble wall\,\footnote{Friction due to splitting radiation \cite{Bodeker:2017cim,Hoche:2020ysm,Azatov:2020ufh,Gouttenoire:2021kjv} is subdominant as long as the gauge coupling $g' \lesssim {\cal O} (0.1)$ and the wall Lorentz factor $\gamma_w = {\cal O} (1)$, as is the case in our scenarios.} due to a full thermal distribution of particles crossing into the bubble and becoming massive\, is $\mathcal{P}_{\text{max}}\approx \frac{1}{24}m^2 T^2$ \cite{Bodeker:2009qy}. 
For $m>\sqrt{g_*^D\alpha} \,T$ (we dropped an $\mathcal{O}(1)$ prefactor for simplicity), we have $\Delta V < {\cal P}_{\rm max}$, and the released energy is entirely absorbed by an appropriate fraction of the $X$ population crossing into the bubble and becoming massive, resulting in a steady-state, terminal bubble wall velocity.
\begin{figure}
\centering
\includegraphics[width=0.4\textwidth]{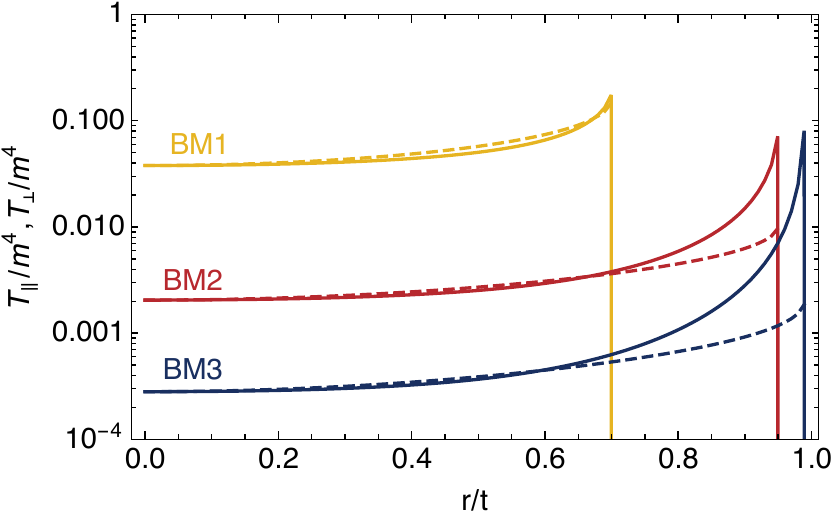}
\caption{
Energy-momentum profile of particles inside an expanding bubble.
The solid (dashed) curves denote $T_{\parallel}$ ($T_{\perp} $). 
}
\label{fig:profile}
\end{figure}

For our numerical studies, we focus on three benchmark (BM) cases (which satisfy $\gamma T > m$):
\begin{equation}
	(m/T,v_w)= \begin{Bmatrix} {\rm BM1} && {\rm BM2} && {\rm BM3}\\
	(1,  0.7)&& (2,0.95)&& (3,0.99)
	\end{Bmatrix}. \label{eq:bms}
\end{equation} 
Larger values of $m/T$ are also possible, with a smaller fraction of the $X$ population entering the bubbles while the majority gets reflected; we focus on $m/T\sim 1$ purely for convenience, as this does not require keeping track of the reflected population. 

Due to the energy transfer from the bubble walls to the particles, the massive $X$ particles in the broken phase gain momenta in the direction of wall propagation, forming extended shells that trail the walls and expand outwards. Fig.\,\ref{fig:profile} shows the distribution of energy-momentum $T_{ij}$ of particles within a bubble for the BM cases (see Supplemental Material for details of the computation), with $T_\parallel = T_{xx}$ and $T_\perp = T_{yy} = T_{zz}$ for wall motion in the $x$ direction. The profiles are found to be self-similar (depending only on $r / t$, the time-dependent bubble radius divided by the time since nucleation), with distributions more sharply peaked for higher $m/T$ and $\gamma_w$, as faster walls can drag particles along more strongly. The energy is mostly concentrated in extended shells with thickness comparable to the bubble radius, with a loose tail that extends inwards.

We are interested in scenarios where this population of massive particles, or their decay products -- we will denote the relevant particle by $Y$ -- only have feeble interactions (i.e.\,effectively do not interact) over the timescale of the phase transition. In the broken phase, $Y$ could interact with other particles within a bubble during the expansion phase, or with particles inside other bubbles after collision. In both cases, the condition for $Y$ to be noninteracting during the phase transition is $n_Y  \sigma  R_*< 1\,$.
Here $n_Y$ is the average number density of $Y$ particles ($\sim T^3$ assuming a full thermal distribution), $\sigma$ is the relevant interaction cross section, and $R_*$, the average bubble size at collision, represents the timescale over which the PT completes. If $Y=X$,  there are unavoidable $X-s$ and ($s$-mediated) $X$ self-scattering processes arising from the mass-generating coupling; nevertheless, the above condition can be satisfied with appropriate parameters. Alternatively, if $X$ decays rapidly to other dark sector particles $Y$ in the broken phase, such interactions are trivially avoided. We discuss details of the underlying particle physics model in the Supplemental Material. Since the massive FIPs $Y$ could constitute a significant fraction of the total energy in the Universe, we assume that $Y$ is metastable and decays into SM final states after the PT completes in order to avoid potential constraints from overclosure.

\section{Gravitational wave signals}

We now discuss the GW signal generated in the FIP scenario, drawing comparisons with GWs from the more familiar sound wave source.
Gravitational waves $h_{ij}$ are the transverse-traceless part of the Friedmann-Lema\^itre-Robertson-Walker metric $dx^2 = - dt^2 + a^2 (t) (\delta_{ij} + h_{ij}) dx^i dx^j$, sourced by the energy-momentum tensor $T_{\mu \nu}$ through the wave equation $\Box h_{ij} = 16 \pi G \Lambda_{ij, kl} T_{kl}$, where $\Box$, $G$, and $\Lambda_{ij, kl}$ are the d'Alembertian, the Newtonian constant, and the tensor that projects out the transverse-traceless components, respectively. The GW spectrum at the time of production, $\Omega_{\rm GW}^{*}$, is calculated from the Fourier transform of $T_{\mu \nu}$
\begin{align}
\Omega_{\rm GW}^{*} (k)
&\equiv
\frac{1}{\rho_{\rm tot}}
\frac{d \rho_{\rm GW}}{d \ln k}
\nonumber \\
&\propto
\int \frac{d\hat{k}}{4 \pi}
\Lambda_{ij, kl} (\hat{k})
\left.
T_{ij}^* (\omega, \vec{k})
T_{kl} (\omega, \vec{k})
\right|_{\omega = |\vec{k}|},
\label{eq:W}
\end{align}
where $\rho_{\rm tot}$ is the total energy density of the Universe.
 
To calculate GW production, we develop a novel numerical scheme to calculate the energy-momentum tensor from the superposition of FIPs from multiple bubbles. We treat each spacetime point as a `sprinkler', which emits particles with a definite spectrum when a bubble wall passes through it; the GW spectrum for a given bubble nucleation history is obtained by summing all sprinkler contributions. Further details of the calculations and simulations are provided in the Supplemental Material. We compute GW signals for a stable bosonic FIP particle $X$ that is thermalized in the symmetric phase.\,\footnote{For decay into $Y$ particles, the results are expected to be qualitatively similar if $m_Y\sim m_X$. If $m_Y\ll m_X$, the GW signal can be suppressed due to the $Y$ particles being more dispersed.}

\begin{figure}[t]
\centering
\includegraphics[width=0.15\textwidth]{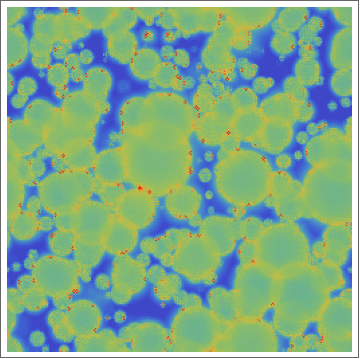}
\includegraphics[width=0.15\textwidth]{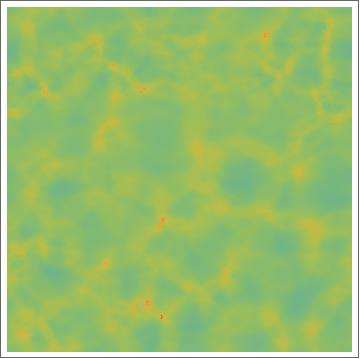}
\includegraphics[width=0.15\textwidth]{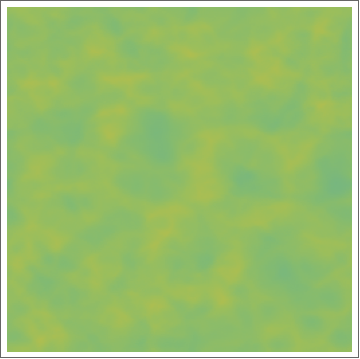}
\includegraphics[width=0.15\textwidth]{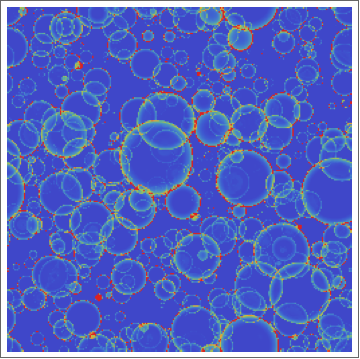}
\includegraphics[width=0.15\textwidth]{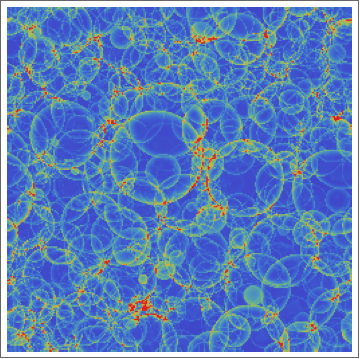}
\includegraphics[width=0.15\textwidth]{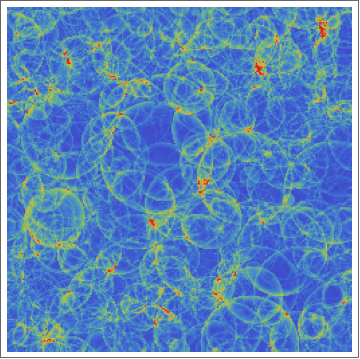}
\caption{
Snapshots of time evolution of $(T_{ij} T_{ij})^{1/4}$ in the FIP scenario (BM1) (top), contrasted with an analogous simulation in the interacting scenario (SWs)~\cite{Jinno:2020eqg} (bottom). These plots are for illustrative purposes only, to highlight the qualitative differences between the two cases. Blue$\to$green$\to$yellow$\to$red (normalized differently for the two cases) denotes increasing $(T_{ij} T_{ij})^{1/4}$; the red dots in the top row are numerical artifacts.
}
\label{fig:time_evolution}
\end{figure}

There are two main differences compared to the SW source.
First, as clearly seen in Fig.~\ref{fig:time_evolution}, the FIP shells extend to the center of the bubble, whereas the sound shells have clear endpoints corresponding to the sound speed $c_s \simeq 1 / \sqrt{3}$ determined by hydrodynamics~\cite{Espinosa:2010hh,Giese:2020rtr,Giese:2020znk,Tenkanen:2022tly}. 
The second concerns the behavior of $T_{ij}$ when the shells cross after bubbles collide.
For the FIPs case, since the particles free-stream without interacting, the total energy is obtained by simply adding the individual contributions from each bubble
\begin{align}
T_{ij} (t, \vec{x})
&=
\sum_{I:~{\rm bubbles}} T_{ij}^{(I)} (t, \vec{x}).
\end{align}
Although the shells cross without interacting and superimpose trivially, the spherical symmetry of each shell  nevertheless gets broken after collision (not visible in the figure): Particles that enter the bubble from the symmetric phase gain mass and get dragged along with the wall, but particles that enter from another bubble are already massive and therefore maintain their inward velocity, resulting in a non-isotropic distribution of particles in the shells.

In contrast, for the interacting fluid (sound shells), the linearlized fluid equation of motion $(\partial_t^2 - c_s^2 \nabla^2) v^{\rm (fluid)}_i = 0$ (neglecting vorticity) implies that the \textit{fluid velocity field}, rather than the energy momentum, superimposes linearly
\begin{align}
v^{\rm (fluid)}_i (t, \vec{x})
&=
\sum_{I:~{\rm bubbles}}
v^{{\rm (fluid)} (I)}_i (t, \vec{x}).
\end{align}
The GW source thus behaves nonlinearly in the superposition of the fluid shells
\begin{align}
T^{\rm (fluid)}_{ij} (t, \vec{x})
&\sim
w (t, \vec{x}) v_i (t, \vec{x}) v_j (t, \vec{x})
+ (\delta_{ij}~{\rm piece}),
\end{align}
where $w$ is the fluid's enthalpy. Consequently, new correlations get imprinted at the scale $k \sim {\rm (shell~thickness)}^{-1}$, allowing for the accumulation of GWs at the same scale long after the collisions take place, enhancing the signal by a factor $\lesssim \beta/H$ \cite{Hindmarsh:2013xza,Hindmarsh:2015qta,Hindmarsh:2017gnf,Cutting:2019zws,Hindmarsh:2016lnk,Hindmarsh:2019phv}. For FIPs, as the particle shell thickness continues to expand as the FIPs propagate, the signal is instead imprinted over a larger range of wavenumbers, resulting in a broader signal.

\begin{figure}[t]
\centering
\includegraphics[width=0.45\textwidth]{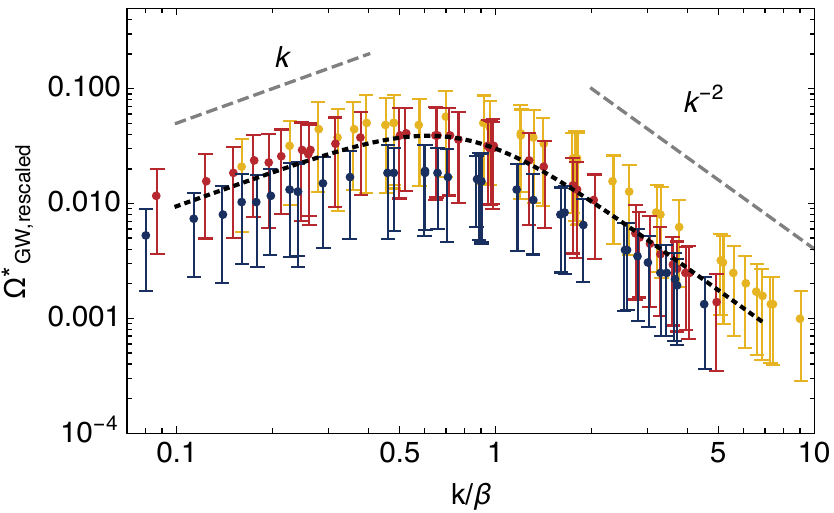}
\caption{Gravitational wave spectrum obtained from simulations with 50 nucleation histories, rescaled as $\Omega^*_{\rm GW}/[(\bar K^{(\rm GW)})^2 (\frac{1}{24}m^2 T^2/\rho_{\rm tot})^2 (H/\beta)^2$]. The dashed black curve shows the broken power law fit $s/(1+s^3)$.
The error bars correspond to the variance associated with the average over propagation directions.
We have checked that the variance due to the different nucleation histories is significantly smaller.
}
\label{fig:Omegaf}
\end{figure}

We evaluate Eq.\,(\ref{eq:W}) to obtain the GW signals from FIPs (see Supplemental Material for details) for the three BM cases (Eq.\,\ref{eq:bms})
and show the result in Fig.~\ref{fig:Omegaf}. The resulting signal (at production) can be parameterized as 
\begin{align}
\Omega_{\rm GW}^*(k)
&\!\sim
\left(\frac{H}{\beta}\right)^2 \left(\frac{\frac{1}{24}m^2 T^2}{\rho_{\rm tot}}\right)^2 \left[\left( \bar K^{(\rm GW)}\right)^2\,\frac{2s}{1+s^{3}}\right],
\label{eq:fit1}
\end{align}
with $s = 0.77 \times k/\beta$. Here, $\frac{1}{24}m^2T^2 \approx \Delta V$, hence the factor $\left(\frac{\frac{1}{24}m^2 T^2}{\rho_{\rm tot}}\right)^2$ represents the characteristic scaling $(\alpha/(1+\alpha))^2$ of GW signals, which is also observed for GW signals from SWs and bubble collisions. The part in the square parenthesis, obtained from a fit to the simulation data, consists of two pieces. 
The spectral \textit{shape} is approximately universal: the spectra peak at $k\sim 0.77\beta$, and scale as $\sim k^1 (k^{-2})$ in the IR (UV). In the far IR, we expect the shape to scale as $\sim k^3$ as correlations are lost beyond a Hubble time; we do not recover this scaling in our simulations as we ignore the expansion of the Universe.

\begin{figure}[t]
\centering
\includegraphics[width=0.45\textwidth]{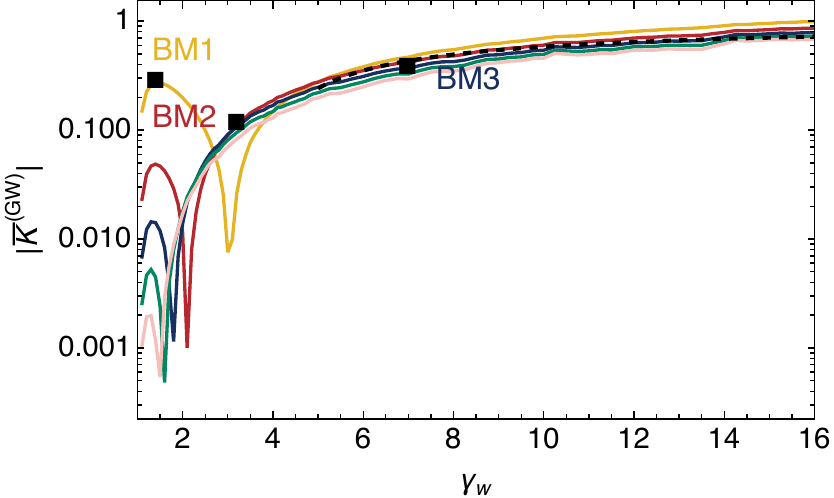}
\caption{
Absolute value of $ \bar{K}^{\rm(GW)} $ as a function of $\gamma_w$ for $m/T = (1,2,3,4,5)$ (top to bottom). The black dashed curve is a numerical fit with formula $\bar K^{\rm(GW)} \sim 1 - 3/\gamma_w^{0.86}$. The black squares represent our BM points. }
\label{fig:fbar}
\end{figure}

The function $\bar K^{(\rm GW)}$ encodes the details of the underlying process, i.e.\,the dependence on $m/T$ and $v_w$ (see Supplemental Material for derivations).
It quantifies the fraction of energy in the FIP distribution that is relevant for GW production.\,\footnote{
$\bar K^{(\rm GW)}$ is analogous to the kinetic energy fraction $K$ for SWs (but also includes projection onto the transverse-traceless modes).
} The bar denotes an average over all propagation directions. We provide numerical values for various choices of $(m/T, \gamma_w)$ in Fig.\,\ref{fig:fbar}. $\bar K^{(\rm GW)}$ approaches universal behavior at large $\gamma_w$.
At small $\gamma_w$, $|\bar K^{(\rm GW)}|$ features a dip at a particular value of $\gamma_w$ for each $m/T$; this is because $\bar K^{(\rm GW)}$ roughly correlates with the average radial velocity of the particle distribution, and for each $m/T$ there exists a $\gamma_w$ for which the particle distribution is approximately static in the plasma frame, making GW production inefficient. 

\begin{figure}[t]
\centering
\includegraphics[width=0.46\textwidth]{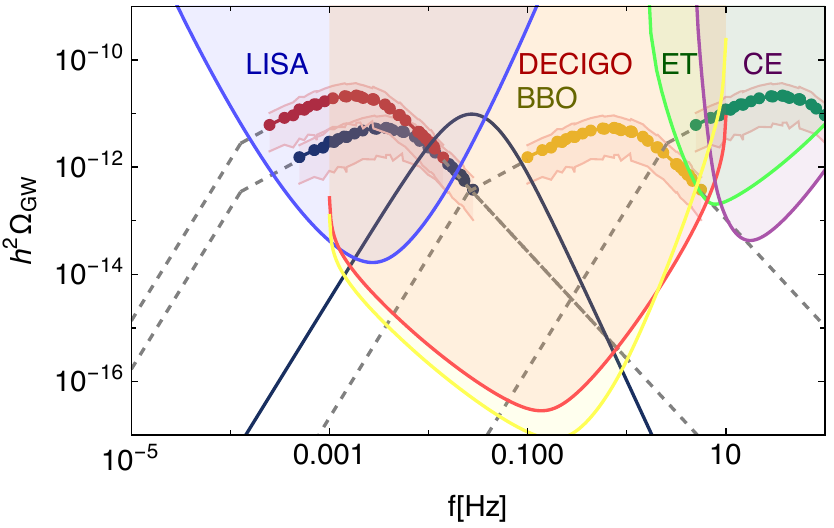}
\caption{
Gravitational wave signals from FIPs with $m/T=3$, $v_w = 0.99$, and $g_*^D=5$ (which determine $\alpha = 0.23$). The different curves correspond to $T=5~{\rm TeV}, ~\beta/H = 100$ (blue), $T=5~{\rm TeV},~\beta/H = 50$ (red), $T=1~{\rm PeV},~\beta/H = 100$ (yellow), and $T=100~{\rm PeV},~\beta/H = 50$ (green). For comparison, we also show a sound wave signal (solid curve), using the same PT parameters as the blue curve, and $\alpha = 0.23$. The power-law integrated sensitivity curves for GW experiments \cite{Harry:2006fi,Kawamura:2006up,Punturo:2010zz,2017arXiv170200786A,Reitze:2019iox} are for 1 year observation time, with signal-to-noise ratio $=1$, obtained from \cite{Schmitz:2020syl}.
}
\label{fig:SignalandExps}
\end{figure}

Rescaling Eq.\,\ref{eq:fit1} by the appropriate redshift factors gives the present day values
\begin{equation} 	
	{ f = 1.2 \times 10^{-6} {\rm Hz} \frac{\beta}{H_*} \frac{T}{100 {\rm GeV}} \left(\frac{g_*^D}{5} \right)^{1/4} \left( \frac{106.75}{g_{\rm SM}} \right)^{1/12} \frac{k}{\beta}},
\end{equation}
and
\begin{equation}
	{\Omega_{\rm GW}^0 = 5.0 \times 10^{-5} \frac{5}{g_*^D} \left(\frac{106.75}{g_{\rm SM}} \right)^{1/3} \Omega_{\rm GW}^*}.
\end{equation}
In Fig.\,\ref{fig:SignalandExps}, we plot the resulting GW signals for a few scenarios\,\footnote{As discussed in \cite{Lewicki:2022nba}, $\alpha$, $v_w$, and $m/T$ are not independent quantities, but are related by details of energy transfer. The parameters chosen here are roughly consistent with the relation found in \cite{Lewicki:2022nba}.} corresponding to PTs at temperatures ranging from $5 {\rm \, TeV}$ to $100 {\rm \, PeV}$ against the power-law integrated sensitivities of various upcoming GW detectors. The signal curves present the results of our simulations (colored dots) extrapolated in the UV with a $k^{-2}$-tail, and with a $k^{1}$-tail in the IR, which breaks into a $k^{3}$-tail at the Hubble scale at the time of the transition. If the FIPs decay soon after the PT completes, we expect the IR component to shut off exponentially, leaving another characteristic imprint on the signal. For comparison, we also include the GW signal predicted from a SW source (blue curve, made with {\tt{PTPlot}}\,\cite{Caprini:2019egz}), with the same PT parameters as the blue FIP curve. 

The peak position of the SW signal is set by the size of the sound shells \cite{Hindmarsh:2017gnf}, whereas the peak of the FIP signal is set by the bubble size, explaining the order of magnitude difference in peak frequency.
The amplitude of the GW signal from FIPs is comparable to the SW signal: despite the $\beta/H$ enhancement of the latter due to the signal accumulating at the same scale, FIPs appear to be more efficient at sourcing GW signals as there is no energy 
loss inefficiency due to their noninteracting nature.  The plot shows that various upcoming space- or ground-based GW detectors can be sensitive to FIP-GW signals produced from such PTs across a broad range of energy scales. Furthermore, the accumulation of the GW signal over a broader range of wavenumbers also results in the FIP signal having a broader peak that scales as $\sim k^1$, which is a distinguishing characteristic of this scenario.

\section{Discussion}

We have discussed a novel source of gravitational waves from a FOPT: feebly interacting particles carrying the dominant fraction of the latent energy released in a phase transition in a dark sector. This provides an interesting and realistic alternative to conventionally studied FOPT GW sources, and we have demonstrated that the GW spectra are qualitatively different: compared to a SW source with similar PT parameters, the FIP-induced signal has comparable amplitude but is broader, and scales differently with frequency. The spectral shape is most similar to that from the bulk flow model \cite{Jinno:2017fby,Konstandin:2017sat}, which models GW production from thin shells; the crucial difference with our setup is the extended FIP shell thickness, which can imprint distinguishable UV features in the signal. We have shown that these novel GW signals could be observable in next-generation GW experiments.\,\footnote{A FIP-induced signal from an MeV scale dark phase transition could also explain the signal recently observed by NANOGrav \cite{NANOGrav:2021flc}.} Such FIP configurations appear to be particularly efficient GW sources, since their noninteracting nature eliminates energy loss inefficiencies, and the GW signals are comparable in strength to those encountered with SW sources. We have provided an analytic formula in Eq.\,\ref{eq:fit1} that (together with Fig.\,\ref{fig:fbar}) can be used to estimate the GW signal from FIPs in specific BSM models and study detection prospects without performing numerical simulations. 

We studied a simplified, idealized scenario where the energy released in the phase transition goes entirely into FIPs, which act as the only source of GWs.
Realistic scenarios likely involve additional features: for instance, a fraction of the particle distribution will get reflected; some fraction of the FIPs may self-scatter, or decay into interacting SM states on the same timescale as the phase transition.
These can produce SWs in the plasma, which will produce additional GWs, while suppressing the FIP signal. Even in such cases, the two contributions can likely be distinguished in an observed signal from their distinct features (see Fig.\,\ref{fig:SignalandExps}). Similarly, the case where FIPs produced from the decay of particles crossing into the bubble are significantly lighter than the parent particle will also change the signal; nevertheless, we expect the spectral features to remain the same. Such aspects merit further study in the future.

\section*{Acknowledgments} 

We thank Iason Baldes, Oliver Gould, Thomas Konstandin, Marek Lewicki, and Pedro Schwaller for helpful comments. This work is supported by the Deutsche Forschungsgemeinschaft under Germany's Excellence Strategy - EXC 2121 Quantum Universe - 390833306.
The work of RJ is supported by the Spanish Ministry for Science and Innovation under grant PID2019-110058GB-C22 and grant SEV-2016-0597 of the Severo Ochoa excellence program.
JvdV is supported by the Dutch Research Council (NWO), under project  number VI.Veni.212.133.
The authors would like to express special thanks to the Mainz Institute for Theoretical Physics (MITP) of the Cluster of Excellence PRISMA*(Project ID 39083149) for its hospitality and support. BS also thanks the Berkeley Center for Theoretical Physics, the Lawrence Berkeley National Laboratory, and the CERN Theory Group for hospitality during the completion of the project.


\onecolumngrid

\appendix

\vskip 0.4in

\begin{center}
\textbf{Supplemental Material}
\end{center}

\section{Particle Physics Frameworks}\label{app:pms}

The GW signals from feebly interacting particles (FIPs) scenario discussed in this paper requires the dominant fraction of the energy released during a FOPT to be transferred to particles that don't interact over the timescale of the phase transition, i.e. satisfy Eq.\,5. In this Supplemental Material, we provide a broad (but not exhaustive) discussion of the particle physics frameworks that could give rise to such setups. 

The scalar $s$, which undergoes the phase transition, can itself serve as the FIP if it gains a large mass by virtue of the FOPT; however, a large mass also, in general, implies a large quartic coupling $\lambda_s$, which leads to efficient self-scattering in the broken phase.
Instead, the role of $X$ can be played by the gauge boson $Z'$ corresponding to the broken symmetry; this requires a sufficiently large gauge coupling $g'$, such that $m_{Z'}=g'\langle s \rangle$ saturates the condition in Eq.\,3. This coupling $g' m_{Z'}sZ'Z'$ can also give rise to scalar-mediated $s$- and $t$-channel $Z'$ self-scattering processes; assuming $\gamma_t T\approx m_{Z'}< m_s\approx \langle s \rangle$, this self-scattering cross section is $\sigma\sim \frac{g'^4}{(4\pi)^2} \frac{m_{Z'}^2}{ m_s^4}$, and the noninteracting condition (Eq.\,5) is
\be
\frac{(4\pi)^2}{g'^6} \frac{{\langle s \rangle}^2}{T\,M_{Pl}}\frac{\beta}{H} > 1\,,
\label{eq:condition}
\ee
where we have used $R_*\approx 1/\beta$, $H\approx T^2/M_{Pl}$. For $\beta/H\approx 100$ and $T\sim \mathcal{O}($TeV), this implies $\langle s\rangle/(g'^2 T) >10^6$. Recalling that Eq.\,3 also requires $g' \,{\langle s\rangle}/T > \mathcal{O}(1)$, satisfying the above condition generally requires $g'>0.01, \,\langle s\rangle > 100\,T$. In addition, the $Z'$ can also scatter with the scalar, but for $m_s\approx \langle s \rangle \gg m_Z, T$ the scalar has a suppressed population in the broken phase, making such scatterings negligible. With this mass hierarchy, inverse decays $Z'Z'\to s$ and resonant contributions to $s$-channel $Z'$ scattering \cite{Frangipane:2021rtf} are also negligible.
While the presence of gauge boson species gives rise to an additional contribution to the friction pressure to the wall, this effect is negligible as long as the gauge coupling and the wall relativistic factor are in the range $g' \lesssim {\cal O} (0.1)$ and $\gamma_w = {\cal O} (1)$, as their ratio goes as ${\cal P}_{\rm LL} / {\cal P}_{\rm LO} \sim g'^2 \gamma_w m_{Z'} T^3 / \frac{1}{24} m_{Z'}^2 T^2 = 24 g'^2 \gamma_w T / m_{Z'}$~\cite{Bodeker:2017cim,Gouttenoire:2021kjv} (here the subscripts LL and LO stand for ``leading log" and ``leading order"). Note that both the LO and LL friction can be negative  \cite{Long:2025qoh,Shakya:2025mdh,Shakya:2025qpi} in some (symmetry-restoring) FOPTs , requiring a more careful treatment of the terminal bubble wall velocity.

Any other particle that gets massive through its coupling to the scalar $s$ will also have similar scattering cross sections mediated by this coupling, and faces similar constraints. Thermally triggered phase transitions generally occur at $T\sim\langle s\rangle$, hence the $\langle s\rangle > 100\,T$ hierarchy likely requires some nontrivial setup, such as supercooled transitions \cite{Konstandin:2011dr,vonHarling:2017yew,Baratella:2018pxi,DelleRose:2019pgi,Fujikura:2019oyi,Ellis:2019oqb,Brdar:2019qut,Baldes:2020kam,Baldes:2021aph}, or transition via quantum tunnelling.\,\footnote{In such scenarios, a SM radiation bath at a higher temperature might be required to avoid potentially problematic vacuum dominated inflationary phases.} We do not pursue the details of such setups further, but simply emphasize the general point that any particle that gets its mass from the phase transition and satisfies Eq.\,3 is likely to self-scatter over the course of the phase transition unless $\langle s\rangle\!>\!100\,T$. 

Another plausible possibility is that particle $X$ (which could be $s$, $Z'$, or some other particle in the dark sector with significant coupling to $s$) decays rapidly into FIPs in the broken phase. 
As a representative case, consider $Z'$ boson decay into a pair of fermions $\psi$ (corresponding to the particle $Y$) in the broken phase, via the interaction $Z'\to\psi\bar{\psi}$, with some effective coupling $\epsilon$. Since the massive $Z'$ particles move in the plasma with velocities comparable to the wall velocity, the decay rate is $\Gamma_{Z'}\approx \frac{\epsilon^2}{8\pi\,\gamma_w}m_{Z'}$. The corresponding decay lifetime is much shorter than the timescale of the phase transition, i.e. $\Gamma_{Z'}\gg\beta\sim 1/R_*$, provided
\begin{align}
&\epsilon^2\gg\frac{8\pi \gamma_w}{m_{Z'}R_*}\approx \frac{8\pi}{\gamma_w}\frac{m_{Z'}}{M_{Pl}}\frac{\beta}{H}\,,
\label{eq:lifetime}
\end{align}
where we have written $R_*\approx 1/\beta$, $H\approx T^2/M_{Pl}$, and approximated $\gamma_w T\sim m_{Z'}$. Note that, in this case, a thermal population of $\psi$ is also likely present in the symmetric phase; these contribute to scattering processes but not the gravitational wave signals if $\psi$ does not gain mass from $\langle s \rangle$. 

Again, $\psi$ self-scattering is mediated by $s$- and $t$-channel $Z'$ exchange processes, with cross section $\sigma\sim \frac{\epsilon^4}{(4\pi)^2} \frac{s}{(s-m_{Z'}^2)^2+\Gamma_{Z'}^2 m_{Z'}^2}$ (for $s$-channel) where $\sqrt{s}=E_{cm}$. The collisions typically occur with $E_{cm}\sim \gamma_w m_Z$, for which $\sigma\sim \frac{\epsilon^4}{(4\pi)^2} \frac{1}{\gamma_w^2 m_{Z'}^2}$ (the $t$-channel cross section is comparable), and the noninteracting condition is\,\footnote{This simple estimate ignores resonant enhancement of the cross section at $E_{cm}=\sqrt{s}\approx m_Z$. The enhancement can be evaluated numerically (see e.g. \cite{Chu:2013jja,Frangipane:2021rtf}), and we have checked that the naive estimate above can be enhanced by a few orders of magnitude for the parameters we consider, but this does not change the subsequent estimates or conclusions. Note that the full resonantly enhanced cross section also automatically incorporates the case of inverse decay $\psi\psi\to Z' (\to \psi\psi)$ when the mediator goes on-shell (see discussions in e.g. \cite{Giudice:2003jh,Belanger:2018ccd,Frangipane:2021rtf}).}
\begin{align}
&T^3 \frac{\epsilon^4}{(4\pi)^2} \frac{1}{\gamma_w^2 m_{Z'}^2} R_* < 1\, ~~~~~\Rightarrow~~~~~\epsilon^4<(4\pi)^2 \gamma_w^3 \frac{m_{Z'}}{M_{Pl}}\frac{\beta}{H}\,.
\label{eq:condition2}
\end{align}
A consistent framework must therefore satisfy Eqs.\,3,\,\ref{eq:lifetime},\,\ref{eq:condition2}. We find that this is possible in a large region of parameter space; for instance, $\alpha\sim0.3,\,\gamma_w\sim3,\, \beta/H\sim 100$, $m_{Z'}\sim$ TeV, and $\epsilon\sim 10^{-6}-10^{-4}$. 

Note that we only considered wall friction due to the particle $X$. The scalar is also present in the bath; however, there are two additional considerations for the scalar: (1) If the scalar mass $m_S$ is much greater than the temperature $T$ at which the transition takes place (as is the case for some of the scenarios we consider, where $m_S\sim \langle s\rangle >100 T$), the number density of the scalars is exponentially suppressed by a factor $e^{-m_S/T}$ compared to the standard thermal abundance, hence their crossing into the bubble would not be a significant source of friction, even though its mass is much larger than the mass of $X$. (2) Unlike the X particle, which we have taken to be the gauge boson of the broken symmetry, which therefore gains its mass entirely from the phase transition, a scalar can already be massive in the symmetric phase. Thus, even if $m_S>m_X$, the mass \textit{gain} for the scalar, $\Delta m_S$, across the phase transition could be smaller, in which case the friction due to scalars crossing into the bubbles would be accordingly weaker. Furthermore, for the case where $Z'\to\psi\psi$, $\psi$ could also provide some friction contribution; however, $\psi$ also does not necessarily have to gain mass from the phase transition, in which case it would not provide any friction at bubble crossing if its mass is the same in both phases. Given these additional subtleties, we chose to be conservative and only consider the friction due to the particle $X$ in this paper.

Alternatively, one could also have scenarios where particles that are not originally in the bath get produced efficiently through the interactions of the bubble wall with the plasma \cite{Azatov:2020ufh,Azatov:2021ifm,Azatov:2021irb,Baldes:2021vyz}, or from the dynamics of the background scalar field \cite{Watkins:1991zt,Falkowski:2012fb,Mansour:2023fwj,Shakya:2023kjf,Giudice:2024tcp,Cataldi:2024pgt}. While such particles have small number densities compared to thermal abundances, they can be far more massive than the scale of the phase transition, and therefore could carry a large fraction of the energy density released in the course of the phase transition.  

In summary, there exist several particle physics frameworks where the energy released in a FOPT could be primarily carried away by FIPs, leading to the production of the GW signals that we have explored in this Letter.

\section{Kinematics of particles entering the bubble}\label{app:kin}

Here, we discuss the relations between kinematic properties of a particle before and after entering a bubble. Using unprimed and primed notation for quantities in the symmetric (false vacuum) and broken (true vacuum) phases, the plasma frame energy $E$, four-momentum $p^\mu$, and velocity $\vec{v}$ of a particle in the two phases are
\begin{eqnarray}
	E = \sqrt {|\vec p|^2} &\equiv p, \qquad E' = \sqrt{ m^2 + |\vec p'|^2} \equiv \sqrt{ m^2 +  p'^2},\qquad\vec v = \vec p/E, \qquad\vec v' = \vec p'/E',\nonumber\\
	\vec p &= p_\parallel \hat n + \vec p_\perp, \qquad \vec p' = p'_\parallel \hat n + \vec p'_\perp, \qquad \text{with} \qquad p_\parallel = \vec p \cdot \hat n,~p'_\parallel = \vec p' \cdot \hat n,
\end{eqnarray}
where $\hat n$ is a unit vector orthogonal to the bubble wall pointing outwards, parallel to the bubble wall motion.
In the wall frame, a particle needs to move towards the bubble, and with sufficient momentum in the $\hat n$-direction to enter the bubble and become massive. These conditions correspond to
\begin{equation}
	p_\parallel - v_w E < 0, \qquad \gamma_w^2(p_\parallel - v_w E)^2 -m^2 > 0.
\end{equation}
We can use energy-momentum conservation (note that momentum is not conserved in the $\hat n$ direction) in the wall frame, then boost to the plasma frame to obtain the following relations between the momenta and energies of a particle across the wall:
\begin{align}
E'
&=
\gamma_w^2 (E - v_w p_\parallel) - \gamma_w v_w \sqrt{\gamma_w^2 (- v_w E + p_\parallel)^2 - m^2}, \label{eq:Epinside}
\\
E
&=
\gamma_w^2 (E' - v_w p_\parallel') - \gamma_w v_w \sqrt{\gamma_w^2 (- v_w E' + p_\parallel')^2 + m^2}, \label{eq:Eoutside}
\\
p'_\parallel
&=
\gamma_w^2 v_w (E - v_w p_\parallel) - \gamma_w \sqrt{\gamma_w^2 (- v_w E + p_\parallel)^2 - m^2},
\\
p_\parallel
&=
\gamma_w^2 v_w (E' - v_w p_\parallel') - \gamma_w \sqrt{\gamma_w^2 (- v_w E' + p_\parallel')^2 + m^2},\label{Eq:pparoutside}
\\
\vec{p}_\perp'
&=
\vec{p}_\perp.\label{eq:pperp}
\end{align}

Consider a coordinate system where a bubble nucleates at the origin at $t = 0$. Given a particle at $(t', \vec x')$ with velocity $v'$ inside the bubble, the time $t_c$ and position $\vec x_c$ at which the particle entered the bubble can be evaluated as
\begin{equation}
	v_w^2 t_c^2 = |\vec x_c|^2, \qquad \vec x_c = \vec x' - \vec v'(t'-t_c),
\end{equation}
which yields
\begin{align}
t_c
&=
\frac{(|\vec{v}'|^2 t' - \vec{v}' \cdot \vec{x}') - \sqrt{(|\vec{v}'|^2 t' - \vec{v}' \cdot \vec{x}')^2 - (|\vec{v}'|^2 - v_w^2) (|\vec{v}'|^2 t'^2 - 2 t' \vec{v}' \cdot \vec{x}' + |\vec{x}'|^2)}}{|\vec{v}'|^2 - v_w^2}.
\end{align}

\section{Energy-momentum tensor of particles in a single bubble}\label{app:Tij}

The energy-momentum tensor $T_{ij}$ of particles inside an isolated, expanding bubble is given by
\begin{equation}
	T_{ij}(t',\vec x') = \int \frac{d^3 p'}{(2\pi)^3} \frac{p'_i p'_j}{E'} g' (t, \vec x', \vec p'),
\end{equation}
where $g'$ is the particle distribution function in the broken phase, which we will determine below. We only consider particles that move towards the bubble wall with sufficient momentum to enter it, ignoring the fraction of the population that gets reflected from the wall. We can write $g'(t, \vec x', \vec p')= j(\vec{p}';\vec{x_c}) g(t, \vec x, \vec p)$, where $g(t, \vec x, \vec p)=\frac{1}{e^{E(p')/T} \pm 1}$ is the standard distribution\footnote{The particles in the symmetric phase are assumed to follow a thermal distribution, i.e. Fermi-Dirac or Bose-Einstein distribution (we use the latter for our studies). The thermal distribution is appropriate for the free-streaming case if the particles were in thermal equilibrium at some earlier time. } (with $E$ determined using Eq.\,\ref{eq:Eoutside}), and $j(\vec{p}';\vec{x_c})$ accounts for any change in the distribution resulting from the particles entering the bubble.
 
There are, in principle, two different cases to consider:
\begin{enumerate}
	\item{Free-streaming (fs) case: the particles don't interact in the symmetric phase.}
	\item{ Thermalized (th) case: the particles interact efficiently in the symmetric phase, retaining a thermal distribution.}
\end{enumerate}

In the free-streaming case, Liouville's theorem states that the phase space distribution function remains constant along the trajectories of the system, which implies $ j_{\rm fs}(\vec{p}';\vec{x_c}) = 1$. Thus
\begin{equation}
		T_{ij}^{\rm (fs)}(t',\vec x') = \int \frac{d^3 p'}{(2\pi)^3} \frac{p'_i p'_j}{E'} \frac{1}{e^{E(p')/T} \pm 1}. \label{eq:TeomGeneral}
\end{equation}
The result can be confirmed by an explicit computation of $g'(t,\vec x', \vec p')$ via
\begin{equation}
	g'(t',\vec x', \vec p') \, dx' \wedge dy' \wedge dz' \wedge dp'_x \wedge dp'_y \wedge dp'_z =g(t,\vec x, \vec p) \, dx \wedge dy \wedge dz \wedge dp_x \wedge dp_y \wedge dp_z, \label{eq:fwedge}
\end{equation}
where the wedge product relates differential phase-space volume before and after entering the bubble. The distribution functions are related by the determinant 	$g'(t',\vec x, \vec p') = \left| \det \left[\frac{\partial (x,y,z,p_x,p_y,p_z)}{\partial(x',y',z',p'_x,p'_y,p'_z)} \right] \right| g(t,\vec x, \vec p),$ but we will not provide the details of the computation here.

In the thermalized case, individual particles cannot be tracked in the symmetric phase due to efficient interactions, and Liouville's theorem cannot be used. We can derive the expression for $j_{\rm th}(\vec{p}';\vec{x_c})$ by considering the difference between the two cases and using Eq.\,\ref{eq:fwedge}. In the thermalized case, particles undergo multiple scatterings, hence every particle is equally likely to enter the bubble. This is in contrast with the free-streaming case, where particles that move towards the wall with higher velocity enter with a greater flux than slower particles. Particles that move away from the bubble wall with a velocity faster than that of the wall cannot enter at all. 

For the free-streaming case, from $\vec x = \vec x_c - \vec v(t_c -t)$, the differential $d \vec x$ at time $t$ is given by
$	d\vec x = d \vec x_c - d\vec v(t_c -t) - \vec v d t_c$. Here $d \vec v$ is proportional to $d \vec p$ and thus drops out of Eq.\,\ref{eq:fwedge} because of the wedge product with $dp_i$. Using
\begin{equation}
	v_w^2 t_c^2 = |\vec x_c|^2 \quad \rightarrow \quad dt_c = \frac{\vec x_c \cdot d \vec x_c}{v_w^2 t_c},
\end{equation}
we obtain
\begin{equation}
	dx \wedge dy \wedge dz \sim \left(1- \frac{\vec v \cdot \vec x_c}{v_w^2 t_c} \right) dx_c \wedge d y_c \wedge dz_c,
\end{equation}
which holds as long as the wedge product with momenta is taken. It can easily be seen that
\begin{equation}
	dx_c \wedge dy_c \wedge d z_c \sim \left(1 - \frac{\vec v' \cdot \vec x_c}{v_w^2 t_c} \right) dx' \wedge dy' \wedge dz',
\end{equation}
so we arrive at
\begin{equation}
	dx \wedge dy \wedge dz \sim \frac{1 - \vec v \cdot \vec x_c/v_w^2 t_c}{1 - \vec v' \cdot \vec x_c/v_w^2 t_c} \, dx' \wedge dy' \wedge dz' \equiv j_{{\rm fs},x}(\vec{p}';\vec{x_c}) \, dx' \wedge dy' \wedge dz' ,
\end{equation}
where we split $j_{{\rm fs}}$ into space- and momentum-pieces $j_{{\rm fs},x}$ and $j_{{\rm fs},p}$. Note that $j_{{\rm fs},x}$ gets exactly cancelled by $j_{{\rm fs},p}$ relating the primed and unprimed momenta, i.e.
\begin{equation}
	 j_{\rm fs}(\vec{p}';\vec{x_c}) \equiv j_{{\rm fs},x}(\vec{p}';\vec{x_c}) \times j_{{\rm fs},p}(\vec{p}';\vec{x_c})=1. \label{eq:jfreefactor}
\end{equation}
In the thermalized case, we can directly relate $d \vec x = d \vec x_c$
because efficient interactions between particles in the symmetric phase ensure that there is no net flux to/from any volume element in front of the bubble.
Therefore 
\begin{equation}
	dx \wedge dy \wedge dz \sim \frac{1 }{1 - \vec v' \cdot \vec x_c/v_w^2 t_c} \, dx' \wedge dy' \wedge dz' \equiv j_{{\rm th},x}(\vec{p}';\vec{x_c}) \, dx' \wedge dy' \wedge dz'.
\end{equation}
The factor relating the momenta is unchanged compared to the free-streaming case, i.e. $j_{{\rm fs},p}=j_{{\rm th},p}$, hence 
\begin{equation}
	j_{\rm th} = j_{\rm fs} \frac{j_{{\rm th},x}}{j_{{\rm fs},x}} = \frac{1}{1-\vec v(\vec{p}') \cdot \vec x_c/v_w^2t_c}.
\end{equation}
Therefore, the expression for the thermalized case is
\begin{equation}
		T_{ij}^{\rm(th)}(t',\vec x') = \int \frac{d^3 p'}{(2\pi)^3} \frac{1}{1-\vec v(\vec{p}') \cdot \vec x_c/v_w^2t_c} \frac{p'_i p'_j}{E'}  \frac{1}{e^{E(p')/T} \pm 1} \, \theta \left[{1 - \frac{\vec v(\vec{p}') \cdot \vec x_c}{v_w^2t_c}}\right],\label{eq:TeomThermalized}
\end{equation}
where $\theta$ is the Heaviside function. 

We use the thermalized case for our studies and simulations. A realistic scenario would lie somewhere in between the two cases; numerically, we find that the resulting GW spectrum is qualitatively similar in both cases.

\section{Computation of the gravitational wave spectrum}\label{app:compGW}

The result from the previous section, while helpful in determining the single-bubble profile before collision, is not useful for calculating the energy-momentum distribution and the resulting GW signal in many-bubble systems:
 to apply Eq.\,\ref{eq:TeomThermalized} to multiple bubbles, for every momentum $\vec{p}'$ one has to know all the possible collision points $\vec{x}_c$ and the corresponding velocities in the symmetric phase $\vec{v} (\vec{p}')$.
While this can be done, we employ a different approach that is simpler and more elegant, which we refer to as the \textit{sprinkler picture}. The main idea (see Fig.\,\ref{fig:sprinkler}) is to treat every spatial coordinate as a ``sprinkler" that gets turned on when a bubble wall passes through it, emitting a particle spectrum into the broken phase in the direction in which the wall passes. The energy-momentum tensor can then simply be obtained by adding the contributions from all the sprinklers. 

\subsection{Sprinkler picture}

Note that Eq.\,\ref{eq:TeomThermalized} can be written as
\begin{align}
T_{ij} (t', \vec{x}')
&=
\sum_{{\rm particle}~p}
\delta^{(3)} \left( \vec{x}' - \vec{x}'^{(p)} (t') \right)
\frac{p'^{(p)}_i p'^{(p)}_j}{E'^{(p)}},
\end{align}
where $\vec{x}'^{(p)} (t')$ describes the trajectory of a single particle $p$. Since it is practically impossible to track the trajectory of each particle throughout the entire history of the system, we start tracking it just before the particle enters the broken phase.
The trajectory inside the bubble is
\begin{align}
\vec{x}'^{(p)} (t')
&=
\vec{x}_c + \vec{v}' (\vec{p}; \vec{x}_c) (t' - t_c (\vec{x}_c)).
\end{align}
\begin{figure}
\begin{center}
\includegraphics[width=0.4\columnwidth]{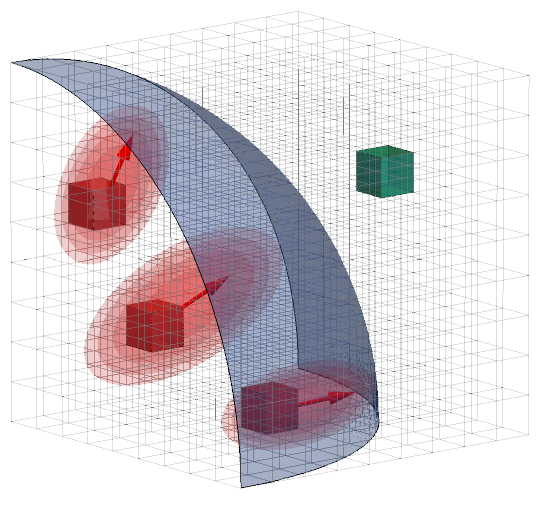}
\caption{\small
Illustration of the sprinkler picture.
The sprinklers in the symmetric/unbroken phase (green cube) are not yet switched on, and contain a thermal distribution of interacting particles. They get switched on (red cubes) when an advancing bubble wall (blue surface) passes through them.
The emission of the free-streaming particles from each sprinkler (red ellipses) is determined only by the sprinkler position $\vec{x}_c$, the collision time $t_c (\vec{x}_c)$, and the direction $\hat{n} (\vec{x}_c)$ (red arrow), and is otherwise universal.
Since each sprinkler contributes to the energy-momentum tensor linearly, one can first calculate the GW emission from the particles from each sprinkler, then sum these contributions, taking into account the $\vec{x}_c$ dependence of each sprinkler.
}
\label{fig:sprinkler}
\end{center}
\end{figure}

The next task is to calculate the flux of particles entering the bubble. We approximate the particles in front of the bubble wall with the thermal distribution $1 / (e^{E (p) / T} \pm 1)$, neglecting the contribution from reflected components. However, not all particles can enter: some particles are prevented from entering because of the wall potential, while (in the free-streaming case) some particles are moving away from the wall with a velocity greater than the wall velocity. We therefore replace the sum over particles with
\begin{align}
\sum_{{\rm particle}~p}
&\to
\int d^3 x_c
\int_{\rm enter} \frac{d^3 p}{(2 \pi)^3}
\frac{w (\vec{p}; \vec{x}_c)}{e^{E / T} \pm 1},
\end{align}
where $w (\vec{p}; \vec{x}_c)$ is a ``wind factor" that accounts for such effects, $\vec{x}_c$ is the position where particles cross the wall, and the label `enter' on the integral denotes that we only include particles that have sufficient momentum to enter the bubble.
The wind factor for the free-streaming case is defined as $w = j_{{\rm fs},x}/j_{{\rm th},x}$, as needed to compensate for the nontrivial dependence of the volume element and match the result of the previous subsection:
\begin{align}
w (\vec{p}; \vec{x}_c)
&=
\left\{
\begin{array}{cl}
1
& \quad : \quad  {\rm thermalized},
\\[0.5cm]
\displaystyle
1 - \frac{\vec{v} (\vec{p}) \cdot \vec{x}_c}{v_w^2 t_c}
=
1 - \frac{\vec{v} (\vec{p}) \cdot \hat{n} (\vec{x}_c)}{v_w}
=
1 - \frac{\vec{p} \cdot \vec{x}_c}{E v_w^2 t_c}
& \quad : \quad {\rm free-streaming},
\label{eq:wind}
\end{array}
\right.
\end{align}
where $\vec{p}$ is obtained from $\vec{p}'$ via Eqs.\,\ref{Eq:pparoutside} and \ref{eq:pperp}.
The energy-momentum tensor can then be written as
\begin{align}
T_{ij} (t', \vec{x}')
&=
\int d^3 x_c
\int_{\rm enter} \frac{d^3 p}{(2 \pi)^3}~
\delta^{(3)} \left[
\vec{x}' - (\vec{x}_c + \vec{v}' (\vec{p}; \vec{x}_c) (t' - t_c (\vec{x}_c)))
\right]
\frac{p_i' (\vec{p}; \vec{x}_c) p_j' (\vec{p}; \vec{x}_c)}{E' (\vec{p}; \vec{x}_c)}
\frac{w (\vec{p}; \vec{x}_c)}{e^{E / T} \pm 1}.
\label{eq:sprinklerTij}
\end{align}
Note that this definition is not restricted to a single bubble but can be applied to a collection of bubbles.\footnote{To recover the single-bubble profile Eq.\,\ref{eq:TeomThermalized}, one can simply integrate out $\vec x_c$ using the $\delta$-function (properly accounting for the $\vec{x}_c$ dependence of the argument) and switching from $\vec{p}$ to $\vec{p}'$ integration, assuming that collision at any $\vec{x}_c$ is triggered by a single bubble.}

In contrast to the expression in the previous section, we now have integrals over the spatial coordinate $\vec{x}_c$ and the momentum $\vec{p}'$.
The sum over $\vec{x}_c$ denotes the sum over sprinklers. Each sprinkler sprays particles when hit by a bubble wall for the first time (see Fig.~\ref{fig:sprinkler}), and is characterized by two quantities:
\begin{itemize}
\item
$t_c (\vec{x}_c)$:
Time when a spatial point $\vec{x}_c$ in the symmetric phase first encounters a wall (= sprinkler turned on) .
\item
$\hat{n} (\vec{x}_c)$:
Direction of wall motion when it passes $\vec{x}_c$.
\end{itemize}
These quantities encode the information about the nucleation history of the collection of bubbles.
Note that the sprinklers are universal except for $t_c (\vec{x}_c)$ and $\hat{n} (\vec{x}_c)$, and, as we will see below, the universal properties of sprinklers are simple to calculate, making the sprinkler approach an efficient way to calculate the GW spectrum from these configurations.

\subsection{Gravitational wave spectrum}

Gravitational waves, the transverse-traceless components of the metric $ds^2 = - dt^2 + a (t)^2 (\delta_{ij} + h_{ij} (t, \vec{x})) dx^i dx^j$, are produced through the linearized wave equation of motion in Fourier space
\begin{align}
\ddot{h}_{ij} (t, \vec{k}) + k^2 h_{ij} (t, \vec{k})
&=
\frac{2}{M_P^2} \Lambda_{ij,kl} (\hat{k}) T_{kl} (t, \vec{k}).
\end{align}
Here $a (t)$ is the scale factor, $\vec{k}$ is the wave vector of the GWs, the dots indicate time derivatives, $M_P \equiv 1/\sqrt{8 \pi G}$ is the reduced Planck mass with $G$ being the Newtonian constant, $\Lambda_{ij,kl} (\hat{k}) \equiv P_{ik} (\hat{k}) P_{jl} (\hat{k}) - P_{ij} (\hat{k}) P_{kl} (\hat{k}) / 2$ with $P_{ij} (\hat{k}) \equiv \delta_{ij} - \hat{k}_i \hat{k}_j$ is the projection tensor, and $T_{ij}$ is the energy-momentum tensor. Note that the contraction with $\Lambda_{ij, kl} (\hat{k})$ drops all components irrelevant for GW production. We neglect cosmic expansion, 
as the phase transition is expected to complete within a small fraction of Hubble time.

The energy density of GWs $\rho_{\rm GW}$ is well-defined once the modes are well inside the horizon, and its logarithmic decomposition can be calculated as~\cite{Weinberg:1972kfs}:
\begin{align}
\Omega_{\rm GW}^* (k)
&\equiv
\frac{1}{\rho_{\rm tot}}
\frac{d \rho_{\rm GW}}{d \ln k}
=
\frac{k^3}{4 \pi^2 \rho_{\rm tot} M_P^2 V}
\int \frac{d\hat{k}}{4 \pi}
(\Lambda_{ij, kl} (\hat{k}) T_{kl} (\omega = k, \vec{k}))^*
(\Lambda_{ij, mn} (\hat{k}) T_{mn} (\omega = k, \vec{k})),
\label{eq:W}
\end{align}
where integration over $d\hat k$ denotes integration over the angular directions and $\rho_{\rm tot}$ and $V$ are the total energy density of the Universe and the volume of the system\,\footnote{
Note that the volume factor $V$ in the denominator cancels out with the implicit $V$-dependence of $T_{ij} (\omega, \vec{k})$, and $\Omega_{\rm GW}^*$ is independent of $V$.
}, respectively.
The formula implies that only the Fourier component with $\omega = k \, (\equiv |\vec{k}|)$ contributes to GWs.\footnote{Here, $\omega$ and $\vec{k}$ refer to broken-phase quantities (since GW production occurs after the shells cross), but we do not put primes on them for notational simplicity.}

To take advantage of the universality of the sprinklers, we will keep $\vec{x}_c$ unintegrated in Eq.\,\ref{eq:sprinklerTij}.
Taking the Fourier transform of Eq.\,\ref{eq:sprinklerTij} and omitting the label `enter' for simplicity, we have
\begin{align}
T_{ij} (\omega = k, \vec{k})
=
&
\int d^3 x_c
\int \frac{d^3 p}{(2 \pi)^3}
\int_{t_c (\vec{x}_c)} d t'~
e^{i k t'}
\int d^3 x'~
e^{- i \vec{k} \cdot \vec{x}'}
\nonumber \\
&\delta^{(3)} \left[
\vec{x}' - (\vec{x}_c + \vec{v}' (\vec{p}; \vec{x}_c) (t' - t_c (\vec{x}_c)))
\right]
\frac{p_i' (\vec{p}; \vec{x}_c) p_j' (\vec{p}; \vec{x}_c)}{E' (\vec{p}; \vec{x}_c)}
\frac{w (\vec{p}; \vec{x}_c)}{e^{E / T} \pm 1}.
\end{align}
We can easily perform the integrals over $\vec{x}'$ and $t'$.
For the latter, we introduce a small imaginary part so that $t' \to \infty$ does not contribute.
To be more precise, after performing the integration over $\vec{x}'$ with the $\delta$-function, we find
\begin{align}
\int_{t_c (\vec{x}_c)}^\infty d t'~
e^{i k t'}
e^{- i \vec{k} \cdot (\vec{x}_c + \vec{v}' (\vec{p}; \vec{x}_c) (t' - t_c (\vec{x}_c)))}
&=
\int_0^\infty d (t' - t_c (\vec{x}_c))~
e^{i (k t_c (\vec{x}_c) - \vec{k} \cdot \vec{x}_c)}
e^{i (k - \vec{k} \cdot \vec{v}' (\vec{p}; \vec{x}_c) + i 0) (t' - t_c (\vec{x}_c))}
\nonumber \\
&=
\frac{i e^{i (k t_c (\vec{x}_c) - \vec{k} \cdot \vec{x}_c)}}{k - \vec{k} \cdot \vec{v}' (\vec{p}; \vec{x}_c)}.
\end{align}
As this does not take cosmic expansion into account, this assumption breaks down for IR modes $k \lesssim H$.
To account for this, we simply change the slope of the GW spectrum at the corresponding wavenumber in our plots to $\Omega_{\rm GW} \propto k^3$~\cite{Caprini:2009fx}.
We still need to perform the integrals over $\vec{x}_c$ and $\vec{p}$. Here, the universality of the sprinklers simplifies the calculation.
Using $v'_i (\vec{p}; \vec{x}_c) = p'_i (\vec{p}; \vec{x}_c) / E' (\vec{p}; \vec{x}_c)$, the projected energy-momentum tensor becomes\,\footnote{
Including the prefactor $\propto k^3$ in Eq.\,\ref{eq:W}, the overall dependence becomes $\Omega_{\rm GW} \propto k^1$ for modes $H \lesssim k \lesssim \beta$, which is characteristic of bulk-flow type sources~\cite{Jinno:2017fby,Konstandin:2017sat}.
}
\begin{align}
\Lambda_{ij, kl} (\hat{k}) T_{kl} (\omega = k, \vec{k})
&=
\int d^3 x_c~
\frac{i e^{i (k t_c (\vec{x}_c) - \vec{k} \cdot \vec{x}_c)}}{k}
\int \frac{d^3 p}{(2 \pi)^3}~
\frac{\Lambda_{ij, kl} (\hat{k}) p_k' (\vec{p}; \vec{x}_c) p_l' (\vec{p}; \vec{x}_c)}{E' (\vec{p}; \vec{x}_c) - \hat{k} \cdot \vec{p}' (\vec{p}; \vec{x}_c)}
\frac{w (\vec{p}; \vec{x}_c)}{e^{E / T} \pm 1}.
\end{align}
Note that there are only two explicit parameters $\hat{k}$ and $\hat{n} (\vec{x}_c)$ in the $\vec{p}$-integrand. 
The dependence of the primed quantities can be found from the relations in Eqs.\,\ref{eq:Epinside}-\ref{eq:pperp}.
Taking the tensorial structure into account, we may write
\begin{align}
\int\! \frac{d^3p}{(2 \pi)^3}\,
\frac{p'_i (\vec{p}; \vec{x}_c) p'_j (\vec{p}; \vec{x}_c)}{E' (\vec{p}; \vec{x}_c) - \vec{p}' (\vec{p}; \vec{x}_c) \cdot \hat{k}}
&\frac{w (\vec{p}; \vec{x}_c)}{e^{E (\vec{p}) / T} \pm 1}
= \nonumber \\
&K_\delta \delta_{ij}
+
K_{kk} \hat{k}_i \hat{k}_j
+
K_{kn + nk} (\hat{k}_i \hat{n}_j (\vec{x}_c) + \hat{n}_i (\vec{x}_c) \hat{k}_j)
+
K_{nn} \hat{n}_i (\vec{x}_c) \hat{n}_j (\vec{x}_c).
\end{align}
Here $K_\delta$, $K_{kk}$, $K_{kn + nk}$, $K_{nn}$ are functions of $\hat{k} \cdot \hat{n} (\vec{x}_c)$ and model parameters, $K_{\cdots}=K_{\cdots}(\hat{k} \cdot \hat{n} (\vec{x}_c); v_w, m, T).$
Thus, for a given set of model parameters, we can first numerically evaluate $K_{\cdots}$ as functions of $\hat{k} \cdot \hat{n}$.
Writing $\hat{k} \cdot \hat{n} (\vec{x}_c) = c$ for simplicity, and contracting with $\delta_{ij}$, $\hat{k}_i \hat{k}_j$, $\hat{k}_i \hat{n}_j + \hat{n}_i \hat{k}_j$, $\hat{n}_i \hat{n}_j$, respectively, we get
\begin{align}
\left(
\begin{array}{c}
\displaystyle\int \frac{d^3p}{(2 \pi)^3}~
\frac{p'^2}{E' - \vec{p}' \cdot \hat{k}}
\frac{w}{e^{E / T} + 1}
\\[0.3cm]
\displaystyle\int \frac{d^3p}{(2 \pi)^3}~
\frac{(\vec{p}' \cdot \hat{k})^2}{E' - \vec{p}' \cdot \hat{k}}
\frac{w}{e^{E / T} + 1}
\\[0.3cm]
\displaystyle\int \frac{d^3p}{(2 \pi)^3}~
\frac{2 (\vec{p}' \cdot \hat{k}) (\vec{p}' \cdot \hat{n})}{E' - \vec{p}' \cdot \hat{k}}
\frac{w}{e^{E / T} + 1}
\\[0.3cm]
\displaystyle\int \frac{d^3p}{(2 \pi)^3}~
\frac{(\vec{p}' \cdot \hat{n})^2}{E' - \vec{p}' \cdot \hat{k}}
\frac{w}{e^{E / T} + 1}
\end{array}
\right)
&=
\left(
\begin{array}{c c c c}
3 & 1 & 2 c & 1
\\
1 & 1 & 2 c & c^2
\\
2 c & 2 c & 2 (1 + c^2) & 2 c
\\
1 & c^2 & 2 c & 1
\end{array}
\right)
\left(
\begin{array}{c}
K_\delta
\\
K_{kk}
\\
K_{kn + nk}
\\
K_{nn}
\end{array}
\right).
\end{align}
Inverting this gives
\begin{align}
\left(
\begin{array}{c}
K_\delta
\\
K_{kk}
\\
K_{kn + nk}
\\
K_{nn}
\end{array}
\right)
&=
\frac{1}{(1 - c^2)^2}
\left(
\begin{array}{c c c c}
(1 - c^2)^2 & - (1 - c^2) & c (1 - c^2) & - (1 - c^2)
\\
- (1 - c^2) & 2 & - 2 c & 1 + c^2
\\
c (1 - c^2) & - 2 c & (1 + 3 c^2) / 2 & - 2 c
\\
- (1 - c^2) & 1 + c^2 & - 2 c & 2
\end{array}
\right)
\left(
\begin{array}{c}
\displaystyle\int \frac{d^3p}{(2 \pi)^3}~
\frac{p'^2}{E' - \vec{p}' \cdot \hat{k}}
\frac{w}{e^{E / T} \pm 1}
\\[0.3cm]
\displaystyle\int \frac{d^3p}{(2 \pi)^3}~
\frac{(\vec{p}' \cdot \hat{k})^2}{E' - \vec{p}' \cdot \hat{k}}
\frac{w}{e^{E / T} \pm 1}
\\[0.3cm]
\displaystyle\int \frac{d^3p}{(2 \pi)^3}~
\frac{2 (\vec{p}' \cdot \hat{k}) (\vec{p}' \cdot \hat{n})}{E' - \vec{p}' \cdot \hat{k}}
\frac{w}{e^{E / T} \pm 1}
\\[0.3cm]
\displaystyle\int \frac{d^3p}{(2 \pi)^3}~
\frac{(\vec{p}' \cdot \hat{n})^2}{E' - \vec{p}' \cdot \hat{k}}
\frac{w}{e^{E / T} \pm 1}
\end{array}
\right).
\end{align}
\begin{figure}
\begin{center}
\includegraphics[width=0.45\columnwidth]{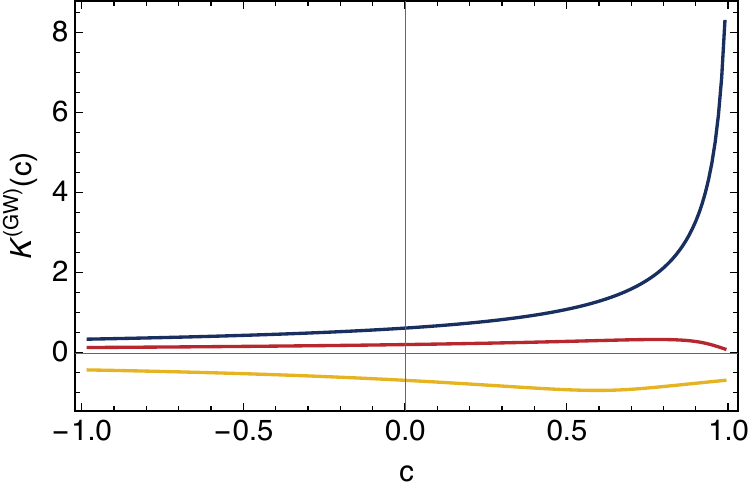}
\caption{\small
$K^{\rm (GW)}$ as a function of $c=\hat{k} \cdot \hat{n} (\vec{x}_c)$ for the benchmark scenarios BM1-BM3 (yellow, red, blue). 
}
\label{fig:fGW2}
\end{center}
\end{figure}
Of these, only the $K_{nn}$ term survives the projection with $\Lambda_{i j, k l} (\hat{k})$ and is relevant for GW signals. We accordingly relabel it as $K^{\rm (GW)}$, and absorb a factor $\sqrt{3/8\pi^2}$ for later convenience:
\begin{equation}
\Lambda_{ij,kl} (\hat k) T_{kl} (\omega = k,\vec k)
=
\frac{1}{24}m^2 T^2 \sqrt{\frac{8\pi^2}{3}}
\int d^3 x_c~
\frac{i e^{i (k t_c (\vec{x}_c) - \vec{k} \cdot \vec{x}_c)}}{k}
\Lambda_{i j, k l} (\hat{k}) \hat{n}_k (\vec{x}_c) \hat{n}_l (\vec{x}_c)
K^{\rm (GW)} \left( \hat{k} \cdot \hat{n} (\vec{x}_c) \right),
\label{eq:LambdaT}
\end{equation}
where
\begin{align}
K^{\rm (GW)} (c)
&= \frac{24}{m^2T^2}\sqrt{\frac{3}{8\pi^2}}
\int \frac{d^3 p~}{(2 \pi)^3}
\frac{2 (\vec{p}' (\vec{p}; \vec{x}_c) \cdot \hat{n} (\vec{x}_c) - c \hat{k} \cdot \vec{p}' (\vec{p}; \vec{x}_c))^2 - (1 - c^2) (p'^2 (\vec{p}; \vec{x}_c) - (\hat{k} \cdot \vec{p}' (\vec{p}; \vec{x}_c))^2)}{(1 - c^2)^2 (E' (\vec{p}; \vec{x}_c) - \hat{k} \cdot \vec{p}' (\vec{p}; \vec{x}_c))}
\frac{w (\vec{p}; \vec{x}_c)}{e^{E / T} \pm 1}.
\label{eq:kgw}
\end{align}
Note that we have chosen to normalize quantities in terms of $\frac{1}{24}m^2 T^2$, which is the vacuum energy density released during the phase transition, to make a clearer connection to the physical energy scale of the setup. The fact that $K^{\rm (GW)} (c)$ depends on $\vec{x}_c$ only through $\hat{n} (\vec{x}_c)$ is the manifestation of the universality of the sprinklers. $K^{\rm (GW)} (c)$ can thus be determined irrespective of the nucleation history. The $\bar{K}^{\rm(GW)}$ in the fit formula for the GW spectrum in Eq.\,10 is precisely this quantity, averaged over its argument $c$. 

In Fig.\,\ref{fig:fGW2}, we plot $K^{\rm (GW)} (c)$  for our benchmark scenarios (for the thermalized case). We see that $K^{\rm (GW)}$ is negative for all values of $c$ for BM1 ($(m/T,v_w)=(1,0.7)$), and (mostly) positive for the other two benchmarks. We find that the value of $K^{\rm (GW)}$ roughly correlates with the average $p'_\parallel$ of the particles inside the bubble. Note that the gravitational wave spectrum does not depend on the sign of $K^{\rm (GW)}$. 

\subsection{Procedure to obtain the gravitational wave spectrum}

\begin{figure}[t!]
\centering
\includegraphics[width=0.6\columnwidth]{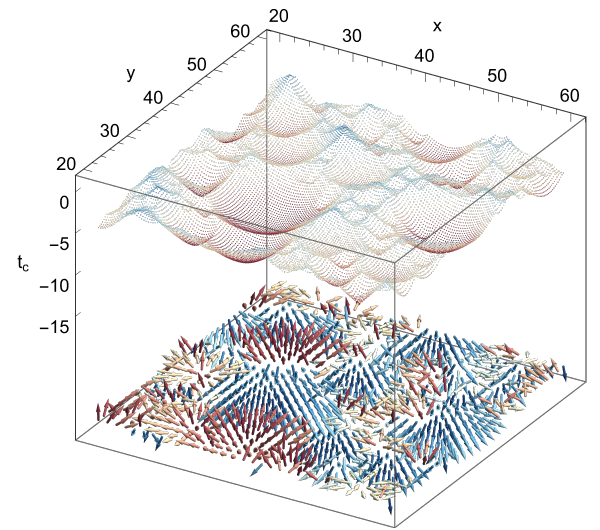}
\caption{\small
Sample collision data for a subset of $x$ and $y$ coordinates along a fixed-$z$ slice $z = 40$.
The data points on the top are the collision time $t_c (\vec{x}_c)$ between the grid point and the first bubble, while the bottom arrows are the direction of the collision $\hat{n} (\vec{x}_c)$ (with red and blue indicating upward and downward collisions, respectively). 
}
\label{fig:tcn}
\end{figure}

In order to obtain the gravitational wave spectrum, we generate a bubble nucleation history in a box of $V = L^3$ with $L = 80 v_w/ \beta$ with the bubble nucleation rate\,\footnote{
The choice of the prefactor $\beta^4$ in the nucleation rate is simply for convenience for numerical simulations: in general $\Gamma (t) = \Gamma_* e^{\beta (t - t_*)}$, and one may define $t = 0$ as the time when $\Gamma (t = 0) = \beta^4$.
} $\Gamma (t) =\beta^4 e^{\beta t}$.
After nucleation, the bubbles expand with wall velocity $v_w$. For each point $x_c$ in the box, we determine the time $t_c$ at which the first bubble wall passes. In Fig.~\ref{fig:tcn} we show the distribution of the first collision time $t_c$ and collision direction $\hat{n}$ (projected onto the 2D plane) in a fixed $z$-slice for a sample simulation.

For given $\vec{k}$, we can now calculate the projected tensor $\Lambda_{ij,kl} (\hat k) T_{kl} (\omega = k,\vec k)$ from Eq.\,\ref{eq:LambdaT}, combining the numerically computed $K^{\rm (GW)} (c)$ (independent of simulations) together with the values of $t_c (\vec{x}_c)$ and $\hat n (\vec{x}_c)$ at each $x_c$ obtained from simulations. The gravitational wave spectrum can then be evaluated as
\begin{equation}
\label{eq:GWsim}
\Omega_{\rm GW}^*(k)
=
\frac{3}{4\pi^2}  \left(\frac{H}{\beta} \right)^2 \left(\frac{m^4}{\rho_{\rm tot}} \right)^2 \left[\left(\frac{k}{\beta} \right)^3\frac{1}{\beta^3 V} \int \frac{d \hat k}{4\pi}
\frac{\beta^4}{m^4}
(\Lambda_{ij, kl} (\hat{k}) T_{kl} (\omega = k, \vec{k}))^*
\frac{\beta^4}{m^4}
(\Lambda_{ij, mn} (\hat{k}) T_{mn} (\omega = k, \vec{k})) \right],
\end{equation}
where the factor in brackets is derived from the simulation. By multiplying the result of the simulation by $(m/T)^4$, the relation to Eq.\,10 becomes clear.

The GW spectra plotted in Fig.\,3 in the main text are obtained from an average of over 50 nucleation history simulations within a box of size $V = L ^3$ with $L = 80 v_w / \beta$ and periodic boundary conditions. The number of bubbles nucleated is $\sim 20,000$, and we take $N^3 = 256^3$ grid points to sample the collision data $t_c (\vec{x}_c)$ and $\hat{n} (\vec{x}_c)$. 
We have confirmed that the obtained results are robust against changes in the grid spacing by varying it by a factor 2.
The code used to simulate the nucleation histories and calculate the GW spectrum is the same one developed in \cite{Jinno:2020eqg, Jinno:2021ury} to study sound wave contributions from phase transitions, and has been extensively tested for issues like sufficient sampling density. 
The code was shown to be sufficient for capturing all physical effects related to sound shells \cite{Jinno:2020eqg}, which are an order of magnitude thinner than the typical bubble size at collision. We have applied the same modules to the physically new framework of shells of feebly interacting particles and the novel sprinkler formalism. The shells of feebly interacting particles are significantly thicker than the sound shells studied in the previous papers. Therefore, the code should be sufficient to capture all physical effects from the FIP shells. 

\subsection{GW data from simulation}
For completeness, we list the data that are displayed in Fig.\,3 in the main text, where $\Omega^*_{\rm GW, rescaled}= \Omega^*_{\rm GW}/[(\bar K^{(\rm GW)})^2 (\frac{1}{24}m^2 T^2/\rho_{\rm tot})^2 (H/\beta)^2]$.

\centering
\begin{tabular}{|c|c|c|c|c|c|}
  \hline
  \multicolumn{2}{|c|}{$m/T = 1$} &\multicolumn{2}{|c|}{$m/T = 2$}&\multicolumn{2}{|c|}{$m/T = 3$}  \\
  \hline
  $k/\beta$ &$\Omega^*_{\rm GW, rescaled}$& $k/\beta$ &$\Omega^*_{\rm GW, rescaled}$ & $k/\beta$ & $\Omega^*_{\rm GW, rescaled}$  \\
  \hline \hline
 0.160 & $0.0211\pm 0.0155$ & 0.0870 & $0.0118\pm 0.00816$ & 0.0801 & $0.00537\pm 0.00364$ \\
 0.227 & $0.0311\pm 0.0214$ & 0.123 & $0.0159 \pm 0.0110$ & 0.113 & $0.00728 \pm 0.00498$ \\
 0.278 & $0.0442\pm 0.0317$ & 0.151 & $0.0181\pm 0.0131$ & 0.139 & $0.00811\pm 0.00606$ \\
 0.320 & $0.0373\pm 0.0287$ & 0.174 & $0.0234\pm 0.0158$ & 0.160 & $0.0105\pm 0.00749$ \\
 0.358 & $0.0451\pm 0.0320$ & 0.195 & $0.0231\pm 0.0171$ & 0.179 & $0.0103\pm 0.00751$ \\
 0.392 & $0.0505\pm 0.0382$ & 0.213 & $0.0261\pm 0.0180$ & 0.196 & $0.0117\pm 0.00813$ \\
 0.453 & $0.0475\pm 0.0351$ & 0.246 & $0.0288\pm 0.0219$ & 0.227 & $0.0131\pm 0.00961$ \\
 0.480 & $0.0504\pm 0.0369$ & 0.261 & $0.0274\pm 0.0209$ & 0.240 & $0.0128\pm 0.0100$ \\
 0.481 & $0.0512\pm 0.0366$ & 0.261 & $0.0289\pm 0.0211$ & 0.240 & $0.0128\pm 0.00931$ \\
 0.577 & $0.0481\pm 0.0339$ & 0.314 & $0.0327\pm 0.0233$ & 0.289 & $0.0148\pm 0.0108$ \\
 0.698 & $0.0564\pm 0.0394$ & 0.379 & $0.0371\pm 0.0253$ & 0.349 & $0.0169\pm 0.0119$ \\
 0.919 & $0.0510\pm 0.0361$ & 0.499 & $0.0395\pm 0.0285$ & 0.460 & $0.0184\pm 0.0136$ \\
 0.960 & $0.0464\pm 0.0318$ & 0.521 & $0.0402\pm 0.0274$ & 0.480 & $0.0182\pm 0.0125$ \\
 1.196 & $0.0416\pm 0.0306$ & 0.649 & $0.0399\pm 0.0291$ & 0.598 & $0.0188\pm 0.0136$ \\
 1.205 & $0.0392\pm 0.0266$ & 0.654 & $0.0398\pm 0.0285$ & 0.602 & $0.0190\pm 0.0131$ \\
 1.318 & $0.0376\pm 0.0275$ & 0.716 & $0.0396\pm 0.0278$ & 0.659 & $0.0182\pm 0.0122$ \\
 1.410 & $0.0328\pm 0.0232$ & 0.766 & $0.0364\pm 0.0263$ & 0.705 & $0.0171\pm 0.0125$ \\
 1.742 & $0.0246\pm 0.0164$ & 0.946 & $0.0318\pm 0.0218$ & 0.871 & $0.0162\pm 0.00993$ \\
 1.777 & $0.0248\pm 0.0177$ & 0.965 & $0.0310\pm 0.0213$ & 0.888 & $0.0162\pm 0.0114$ \\
 1.805 & $0.0240\pm 0.0175$ & 0.980 & $0.0303\pm 0.0213$ & 0.903 & $0.0159\pm 0.0114$ \\
 1.811 & $0.0246\pm 0.0179$ & 0.983 & $0.0320\pm 0.0222$ & 0.905 & $0.0159\pm 0.0113$ \\
 2.343 & $0.0153\pm 0.0108$ & 1.272 & $0.0239\pm 0.0175$ & 1.172 & $0.0120\pm 0.00910$ \\
 2.624 & $0.0126\pm 0.00901$ & 1.425 & $0.0206\pm 0.0145$ & 1.312 & $0.0108\pm 0.00724$ \\
 3.211 & $0.00849\pm 0.00597$ & 1.743 & $0.0144\pm 0.0107$ & 1.605 & $0.00815\pm 0.00564$ \\
 3.306 & $0.00808\pm 0.00585$ & 1.795 & $0.0131\pm 0.00981$ & 1.653 & $0.00826\pm 0.00583$ \\
 3.764 & $0.00630\pm 0.00441$ & 2.044 & $0.0106\pm 0.00728$ & 1.882 & $0.00655\pm 0.00446$ \\
 5.061 & $0.00323\pm 0.00215$ & 2.748 & $0.00558\pm 0.00412$ & 2.530 & $0.00398\pm 0.00282$ \\
 5.181 & $0.00305\pm 0.00215$ & 2.813 & $0.00500\pm 0.00348$ & 2.590 & $0.00400\pm 0.00281$ \\
 5.598 & $0.00247\pm 0.00177$ & 3.039 & $0.00450\pm 0.00323$ & 2.799 & $0.00348\pm 0.00254$ \\
 6.043 & $0.00202\pm 0.00147$ & 3.281 & $0.00366\pm 0.00262$ & 3.021 & $0.00300\pm 0.00216$ \\
 6.607 & $0.00172\pm 0.00124$ & 3.587 & $0.00299\pm 0.00212$ & 3.303 & $0.00252 \pm 0.00181$ \\
 6.830 & $0.00156\pm 0.00112$ & 3.709 & $0.00272\pm 0.00195$ & 3.415 & $0.00251\pm 0.00181$ \\
 7.271 & $0.00135\pm 0.000942$ & 3.947 & $0.00247\pm 0.00169$ & 3.635 & $0.00216\pm 0.00152$ \\
 7.416 & $0.00135\pm 0.000953$ & 4.027 & $0.00245\pm 0.00179$ & 3.708 & $0.00194\pm 0.00136$ \\
 9.057 & $0.00100\pm 0.000719$ & 4.917 & $0.00140\pm 0.00105$ & 4.528 & $0.00133\pm 0.000965$ \\
  \hline
\end{tabular}

\bibliography{GWheavyparticles}{}

\end{document}